\documentclass[conference]{IEEEtran}
\IEEEoverridecommandlockouts
\usepackage{cite}
\usepackage{amsmath,amssymb,amsfonts}
\usepackage{graphicx}
\usepackage{textcomp}
\usepackage{xcolor}
\usepackage{verbatim}
\usepackage{CJK}
\usepackage[top=2cm, bottom=2cm, left=2cm, right=2cm]{geometry}
\usepackage{algorithm}
\usepackage{algorithmicx}
\usepackage{algpseudocode}
\usepackage{booktabs}
\usepackage{threeparttable}
\usepackage{float}

\def\BibTeX{{\rm B\kern-.05em{\sc i\kern-.025em b}\kern-.08em
    T\kern-.1667em\lower.7ex\hbox{E}\kern-.125emX}}
\begin{document}

\title{A Distributed SGD Algorithm with Global Sketching for Deep Learning Training Acceleration\\

\thanks{* Corresponding Author: diaoboyu2012@ict.ac.cn}
}
% 修改了邮箱 增加了Abstract、Conclusion、实验部分，修改了部分实验图片。
\author{\IEEEauthorblockN{Lingfei Dai}
\IEEEauthorblockA{\textit{Institute of Computing Technology} \\
\textit{Chinese Academy of Sciences}\\
Beijing, China \\
dailingfei20s@ict.ac.cn}
\and
\IEEEauthorblockN{Boyu Diao*}
\IEEEauthorblockA{\textit{Institute of Computing Technology} \\
\textit{Chinese Academy of Sciences}\\
Beijing, China \\
diaoboyu2012@ict.ac.cn}
\and
\IEEEauthorblockN{Chao Li}
\IEEEauthorblockA{\textit{Institute of Computing Technology} \\
\textit{Chinese Academy of Sciences}\\
Beijing, China \\
lichao@ict.ac.cn}
\and
\IEEEauthorblockN{Yongjun Xu}
\IEEEauthorblockA{\textit{Institute of Computing Technology} \\
\textit{Chinese Academy of Sciences}\\
Beijing, China \\
xyj@ict.ac.cn}
}

\maketitle

%%% 全文修改
% 1 所有top-k的写法要统一 Done
% 2 所有gs-SGD 要统一，第一次全称，后面全用简称 Done
% 3 gs-SGD  global-sketching SGD，这个说法要全文统一 Done 

% 4 参考gtopk，重新写一下contribution部分 Done

% 5 我们的整体思路就是，topk的方法基于ps，有单点瓶颈问题，但是global之后又有了收敛性降低的问题，然后我们的方法在global的框架下，（采用Count Sketch数据结构减少传输时的梯度损失，）保证了高收敛性，这是我们的核心创新。

% 6 related work收尾部分，我已写好提纲，你来完成最后一段，前后思路统一。 Done

% 7 sketch之后，空间复杂度是多少，vecotor长度是d，复杂度是O(d）？现在文章里还有很多事klogP，k是不是不准确
% 7: O(logd) d is the number of model parameters
% 8 sketch 后的空间复杂度； 我们的global 的通信复杂度；我们翻译出top-k 一个计算复杂度
% 8:空间复杂度O(logd)，通信复杂度O(logd*logP)?

% 9 实验再理一遍，结论与intro统一，全文思路统一
% 10 整体过一遍

%% 全文修改0801
% gTop-K的写法和说法，全文还有一些不统一
% 文中有一些  reduce training time的说法，改成accelerate training的相关说法，和题目统一
% logP  有的没有加2底，查一下，都加上吧

\begin{abstract}
% 随着计算机算力的提升，GPU集群中各机器之间的通信带宽成了限制大规模深度神经网络训练的瓶颈。数据并行的分布式同步随机梯度下降算法已经被广泛用于大规模集群训练场景，但这也对集群中机器之间的通信有着很高的要求以迭代交换梯度。截止目前，已经有许多梯度稀疏化技术，特别是top-K稀疏化技术，被提出用来减少节点之间的通信开销。最近，gtop-k稀疏化scheme被提出，降低通信复杂度从O(kp)到O(logP)，其中k是每个工人选择的梯度的数量，P是工人的数量，这大大提升了top-K算法的可扩展性。然而，gtop-k算法可以在训练模型时的收敛效率上被进一步提升。我们提出了一种在gtop-k基础上，加以运用Count Sketch对梯度进行压缩的方法，能够在部分模型上得到更好的收敛效率，并同时具有O(logP)的通信复杂度，称呼为Global-Sketch S-SGD。我们在4-GPU集群上进行了实验验证，我们的方法具有更好的收敛效率的同时，还相较于gTop-K提升了1.3~3.1×的吞吐量，以及相较于original sketched-SGD提升了1.1~1.2×的吞吐量。

% 缺少一句核心创新的话
% 加上 在global的框架下，采用Count Sketch（或sketching的方法）减少梯度在传输过程中的损失，提高了模型收敛性。
Distributed training is an effective way to accelerate the training process of large-scale deep learning models. However, the parameter exchange and synchronization of distributed stochastic gradient descent introduce a large amount of communication overhead. Gradient compression is an effective method to reduce communication overhead. In synchronization SGD compression methods, many Top-$k$ sparsification based gradient compression methods have been proposed to reduce the communication. However, the centralized method based on the parameter servers has the single point of failure problem and limited scalability, while the decentralized method with global parameter exchanging may reduce the convergence rate of training. In contrast with Top-$k$ based methods, we proposed a gradient compression method with globe gradient vector sketching, which uses the $Count-Sketch$ structure to store the gradients to reduce the loss of the accuracy in the training process, named global-sketching SGD (gs-SGD). The gs-SGD has better convergence efficiency on deep learning models and a communication complexity of O($\log d*\log P$), where $d$ is the number of model parameters and $P$ is the number of workers. We conducted experiments on GPU clusters to verify that our method has better convergence efficiency than global Top-$k$ and Sketching-based methods. In addition, gs-SGD achieves 1.3-3.1× higher throughput compared with gTop-$k$, and 1.1-1.2× higher throughput compared with original Sketched-SGD.

\end{abstract}


\section{Introduction}

%para1 介绍深度神经网络及其应用
% 深度学习近几年得到了广泛的应用，特别是在CV，NLP等应用领域
% 深度学习模型的训练需要大量的数据集
% 随着深度学习模型层数的增加，模型的参数规模增长快速，使得模型的训练变得很慢
In recent years, deep learning has been widely used in various fields, especially in Computer Vision, Natural Language Processing, and Speech Recognition\cite{LeCunBH15}\cite{GuptaVDN20}\cite{MaquedaL0GS18}. Most deep learning networks have achieved state-of-art performance across various domains\cite{Simonyan2015VeryDC}\cite{Szegedy2016RethinkingTI}\cite{Redmon2016YouOL}. Stochastic gradient descent (SGD) and its various variants\cite{Goyal2017AccurateLM} are the main methods for training deep learning networks. However, the training process of deep learning models requires large-scale datasets, like ImageNet\cite{Russakovsky2015ImageNetLS}. As the networks become deeper, the parameters of the model grow quickly results in a longer training time, even taking days to weeks\cite{He2016DeepRL}.

%para2 介绍深度学习网络的训练的硬件（GPU，TPU，MLU）
%为提升深度学习模型的训练效率，
%引出算力需求的大幅提升，分布式硬件的需求，网络带宽的问题，导致了分布式难以大规模训练
%考虑配图
% 通过分布式训练提升训练的时效性已经成为广泛采用的训练方法
% 无论是基于参数服务器的同步训练，还是异步训练，都需要传递大量的梯度参数信息
% 相比于计算的时间，数据传输的时间有数量级的延迟差距，特别是对于需要大量参数交互的网络训练，例如RNN，如何降低参数信息的网络通信开销，是提升分布式训练性能的核心研究点
To improve the training efficiency of deep learning models, high-performance acceleration hardware, like NVIDIA A100\cite{Choquette2020NVIDIAAG}, Google TPU\cite{Jouppi2017IndatacenterPA} and Huawei Ascend910\cite{huawei910}, are applied to reduce training time. In addition, parallel training through multiple acceleration nodes is still one of the most effective ways to reduce training time\cite{mlcommons} significantly. Whether it is synchronous training or asynchronous training, a large number of gradient parameters need to be transmitted between servers. Compared with the computing time, the data transmission time has a delay of orders of magnitude, especially for training models that require a large number of parameter interactions, such as RNN\cite{lstm_nips2015}. The network communication overhead of parameter information is the fatal bottleneck to the performance of distributed training\cite{LiASY14}.

%para3 引出一些分布式学习的方法，并行的一些方法，mini-batch等（参考综述论文），初步指出其问题
%para3首先说明，深度学习网络的并行训练包括数据并行和模型并行，简述其区别和联系，限定本文讨论的范围是数据并行。 数据并行的方案，思路，主要方法等。

In distributed deep learning, data parallelism, model parallelism, and hybrid parallelism\cite{MLSYS2019_c74d97b0} are three commonly used methods. Data parallelism randomly divides the entire datasets into several parts, then dispatches them to different nodes. Each node maintains a complete model replica with local parameters. Model parallelism split the model into several parts delivered on different nodes. Hybrid parallelism is a combination of data and model parallelism. In general, data parallelism is suitable for large training sets, and model parallelism is ideal for large-scale models. The distributed training discussed in this paper is mainly limited to data parallelism. 

% Other methods and problem.
% 加reference
% 最近，许多梯度稀疏化、量化以及压缩方法被提出来显著减小梯度大小以降低分布式训练中工人节点之间的通信开销，并且对模型的收敛率影响不大。在这些梯度压缩算法中，基于streaming算法的Sketching稀疏化方法是比较新颖的， which allows us to recover favorable convergence guarantees of vanilla SGD. 但这种方法的通信复杂度较高。其他通信复杂度较低的梯度稀疏化方法，例如gTop-K，在部分模型训练时的收敛效率仍有进步空间。
% 这一段，还可以再把问题说的明确一些，现有方法的思路还不是太清晰
% 先说DGC的topk方法，再说sketched的方法，以上两个方法有单点瓶颈问题，再说gtop-k的方法，收敛性弱，然后引出我们的方法
%
Recently, many gradient compression methods, including sparsification and quantization \cite{Goyal2017AccurateLM}\cite{hubara2017quantized}\cite{wang2019stochastic}\cite{you2018imagenet}\cite{jia2018highly} have been proposed to reduce the communication overhead between workers in training large-scale DNNs by significantly reducing the size of the exchanged gradients. The most representative one is the Top-$k$ algorithm, which significantly reduces the communication overhead in the distributed training\cite{Lin2018DeepGC}. In addition, the Sketching sparsification method \cite{Ivkin2019CommunicationefficientDS} based on the streaming algorithm is relatively novel, which can recover favorable convergence guarantees of vanilla SGD. But these two methods use the centralized method based on the parameter server (PS), which has the single point of failure problem and limited scalability. However, other methods with the decentralized method, such as gTop-$k$ \cite{Shi2019ADS}, still have room for improvement in convergence efficiency when training deep learning models. 
%%%%%%%%%%%%%%%%%%%%%%%%%%%%%%%%%%%
% 这一段，还可以再把问题说的明确一些，现有方法的思路还不是太清晰
%para4 本文的贡献和思路
% 为了克服以上问题，在这篇论文中，我们提出采用Count Sketch数据结构对工人节点的梯度进行压缩和存储，称之为sketch。然后使用树结构的allreduce算法对sketches进行聚合，这样可以使得通信复杂度从O（kP）减少到O（klogP），同时不影响模型的收敛率。
% In this paper, our solution and contributions.
% gs-SGD 的方法，全分布式的特性，有更好的扩展性，并且通过sketch数据结构进行压缩，有更好的训练收敛率和
% 简述方法和方法达到的效果，然后总结贡献
%To overcome the above problems, in this paper, we propose a decentralized gradient compression method with global Sketching, gs-SGD. 
%%%修改
% 把communication complexity，和topk sketch 分别都比较一下，说明一下d和p的含义是什么。
In this paper, we use the Count Sketch to compress the gradient on the local workers, which is called a sketch. Then we first propose the tree structure All-Reduce algorithm to aggregate the sketches from workers. Compared with gTop-$k$ and Sketched-SGD, we reduce the communication complexity from O($k\log P$) and O($\log d*P$) to O($\log d*\log P$) respectively without affecting the convergence rate of the model. The gs-SGD achieves 1.3-3.1× acceleration effect compared with gTop-$k$, and 1.1-1.2× speedup than original Sketched-SGD. Our contributions are summarized as follows:
\begin{itemize}
	\item[$\bullet$] We propose a global sketch sparsification algorithm on distributed synchronous SGD, called global-sketching SGD, to accelerate distributed training of deep neural networks.
	\item[$\bullet$] We implement the proposed gs-SGD atop popular framework PyTorch and Horovod, and evaluate the training efficiency on a 4-GPU cluster interconnected with 1Gbps Ethernet. Experiments result shows that our algorithm has significantly improved the convergence efficiency of training various CNNs, and reduces the communication time ratio among workers.
	\item[$\bullet$] gs-SGD achieves effectively improved training efficiency on the real-world applications with various CNNs under low-bandwidth networks (e.g., 1 Gig-Ethernet).
\end{itemize}

%para5 本文的组织结构
The rest of the paper is organized as follows. We introduce some related work in Section \uppercase\expandafter{\romannumeral2}, and then present our gs-SGD algorithm in Section \uppercase\expandafter{\romannumeral3}. Experimental results and discussions are presented in Section \uppercase\expandafter{\romannumeral4}. Finally, we conclude the paper in Section \uppercase\expandafter{\romannumeral5}.
\section{Related Work}

% para1 整体的思路介绍，从压缩方法上包括稀疏 量化和一些其他线性变换，sketched
% 数据并行的分布式SGD方法主要包括同步模式和异步模式，本文重点讨论同步模式下的分布式梯度压缩训练方法，按压缩方法可以分为稀疏方法和量化方法，按梯度的更新方式，又可以分为集中模式和分布式模式
In general, data-parallel distributed SGD methods are divided into two categories: synchronous mode and asynchronous mode. This paper focuses on the gradient compression method in synchronous mode. In terms of gradient compression, sparsification and quantization are the two main methods\cite{Alistarh2018TheCO}\cite{2019AConvergence}\cite{2018nips_Quantization}. Both centralized and distributed methods have been applied in different scenario\cite{Vogels2019PowerSGDPL}\cite{Wang2019ScalableDD}\cite{Shi2020CommunicationEfficientDD}.

% para2 量化的方法，举例说明几个，然后说明其问题
% 量化方法的主要思路通过对梯度低精度的表示来实现空间占用率的降低
% one-bit的方法 Seide20141bitSG
% signSGD
% adaptive SGD  Faghri2020AdaptiveGQ
The quantization method achieves a reduction in space occupancy by representing the gradient with a low-precision value. In 1-bit SGD \cite{Seide20141bitSG}, gradient quantization technology is used to reduce the communication overhead of distributed neural network training for the first time. In addition, the method also proposed an error feedback mechanism, which can alleviate the negative impact of gradient quantization and reduce the loss of model accuracy. signSGD\cite{Bernstein2018signSGDCO} combined quantization with gradient direction. The paper demonstrated that it is the direction rather than the value that is significant for gradients. Therefore, even if only the sign of the gradient is kept to update the model, the training process can also converge. In some cases, this can reduce the noise of the gradient and make the convergence time shorter. A recent paper, based on previous work, realizes an adaptive quantization method\cite{Faghri2020AdaptiveGQ}. In General, the value of the gradient is represented by a 32-bit variable. Therefore, there is an upper limit on the ratio of communication reduction for the quantification method, which is up to 32 times.

% para3 稀疏的方法，举例说明，总结其问题
% 相比之下，稀疏的方法可能达到几百倍的梯度压缩效果
% 从阈值的方法说起，再说到top-k的方法，提到top-k方法的几个不同之处（看一些论文）
% 说error-feed back如何降低误差，看一些戴凌飞给的材料
% Aji_2017
% 这些稀疏的都是集中式的解决方法，存在单点的通信瓶颈问题，特别是在严酷的网络通信环境下
% 给出global的论文
In contrast, the sparse method may achieve a gradient compression of hundreds of times\cite{Lin2018DeepGC}. The sparse methods reduce the space occupation by filtering some of the gradients, then discarding the remaining gradients. Filtering rules can be based on thresholds. For different models, how to determine the threshold to filter the gradient is a difficult problem. Therefore, some adaptive methods have also been proposed\cite{Chen2018AdaCompA}. Besides thresholds, Top-$k$ compression is a widely used sparse method\cite{Aji_2017}\cite{Lin2018DeepGC}. DGD (Distributed Gradient Descent)\cite{Aji_2017} is an exploratory work. The training speed can be significantly improved by replacing dense updates in distributed SGD with sparse updates. It got 49\% speedup on MNIST. DGC (Deep Gradient Compression)\cite{Lin2018DeepGC} applies momentum correction and local gradient clipping methods on top of gradient sparseness and achieves over 600 times compression. These sparse methods are mostly centralized, so there is a single point of communication bottleneck problem, especially in the harsh network communication environment. Although some methods have tried distributed gradient synchronization methods\cite{Shi2019ADS}, they still have shortcomings in terms of convergence.

% para4 
% 混合的方法  在量化和稀疏方法的基础上，混合
Beyond quantization and sparse methods, some vector mapping methods and data structures, like Count-Sketch\cite{Charikar2002FindingFI}, are applied to gradient compression\cite{Ivkin2019CommunicationefficientDS}. The sketching data structure is mostly linear, both momentum and error accumulation can be carried out within the data. In the centralized topology, the error and momentum can be transmitted to the central node to eliminate errors and accelerate convergence\cite{Rothchild2020FetchSGDCF}. This feature can also be applied to distributed gradient synchronization topology to reduce the convergence attenuation caused by gradient sparse. 
However, related research is rare.

%para5 本文，总结上文，结合Introduction的，再次强调一下我们的贡献和工作
%para5 本文要解决的是什么问题，和大概思路，与Intro统一
Prior work has proposed applying sketching to compress the gradients on the workers in the distributed training. However, the communication complexity of these methods is high, which leads to large communication overhead. The AllReduce method used in the recently proposed gTop-$k$ effectively reduces the communication complexity of sparsification methods, but its convergence efficiency is still insufficient when training the model. The gs-SGD proposed in this paper has the same low communication complexity as gTop-$K$ and has higher convergence efficiency. The gs-SGD scheme effectively speeds up the training efficiency of various real-world applications of CNNs on GPU clusters under low-bandwidth networks (e.g., 1 Gig-Ethernet).
\section{Methods}
\subsection{SGD with Global Sketching}

% para1 分布式通信的整体介绍，简单一段  all-reduce
% 同步的分布式SGD算法一般分为基于PS的集中式算法和无PS的去中心化算法，
% 在去中心化的场景下，SGD的参数交换和合并是一种典型的all-reduce collective communications操作
% all-reduce
%介绍all-reduce的过程
In general, synchronous distributed SGD algorithms are divided into centralized algorithms based on PSs and decentralized algorithms without PSs. In a decentralized scenario, the exchange and merging process of parameters is a typical all-reduce collective communications operation\cite{Chan2007CollectiveCT}.
Usually, we use the recursive halving and doubling method \cite{Thakur2005OptimizationOC} to all-reduce parameters among clusters. Recursive halving and doubling process is essentially constructing a binary tree to reduce all global parameters, then backtrack the results of the recursion to all leaf nodes along the binary tree. 

% count-sketch的介绍，why use count-sketch？
Various Top-$k$ compression techniques are widely used in gradient compression methods. However, when the compression ratio reaches a higher value, the Top-$k$ method significantly attenuates convergence. In addition, the Top-$k$ method must exchange both the gradient value $j$ and their coordinates $i$  when synchronizing to ensure the correct execution of the reduction, which doubled the amount of communication. 

% In contrast with Top-$k$ compression, there is a data structure named $Count-Sketch$. $Count-Sketch$ can compress a gradient vector $g$ into $S(g)$ with a size of $O(\frac{1}{\epsilon}\log d)$, where $\epsilon$ is a parameter to select top-$k$ largest gradient values, and $d$ is the size of $g$.
% 一段简要介绍sketch的论文，一段简要介绍为什么能线性结合的论文
%%%修改

% 这一部分和上一段有重复的内容，合并一下
% Sketching was proposed early in the streaming data process problems\cite{muthukrishnan2005data}. In gs-SGD, we only focus on one variant of Sketch, $Count-Sketch$ \cite{Charikar2002FindingFI}.
In contrast with Top-$k$ compression, there is a data structure named $Count-Sketch$. $Count-Sketch$ can compress a gradient vector $g$ into $S(g)$ with a size of $O(\frac{1}{\epsilon}\log d)$, where $\epsilon$ is a parameter to select top-$k$ largest gradient values, and $d$ is the size of $g$. $Count-Sketch$ finds all ($\alpha$ ,${\ell}_2$)-heavy coordinates and approximates their values with error $\pm\varepsilon \|g\|_2$. It does so with a memory footprint of $O(\frac{1}{\varepsilon^2\alpha^2}\log d)$. In addition, $Count-Sketch$ gained popularity in distributed systems primarily due to the merge ability property: given a sketch $S(a)$ computed on the input vector $a$ and a sketch $S(b)$ computed on input $b$, there exists a function $S(a+b)$. Note that sketching the entire vector can be rewritten as a linear operation $S(a)=C a$,  and therefore $S(a+b)=S(a)+S(b)$. We take advantage of this crucial property in gs-SGD to aggregate sketches among workers.

% 算法介绍
In gs-SGD, each worker transmits a sketch instead of its gradient. Each worker needs to compress its gradient into a $Count-Sketch$ structure locally. As shown in Lines 8 to 23 of Algorithm 1, in each iteration, two workers whose ranks are separated from each other by the value of the current iteration are taken in turn, and the latter transmits the local sketch to the former. After iterations are complete, we obtain the summed sketch on the first worker. And then, we recover the top-$k$ largest gradient elements by magnitude from the summed sketch and use the exact value of the top-$k$ to update the weight. This algorithm for recovering top-$k$ elements from a summed sketch is summarised in Algorithm 2\cite{Ivkin2019CommunicationefficientDS}.

% 分布式过程的介绍
\begin{figure}
	\includegraphics[width=0.5\textwidth]{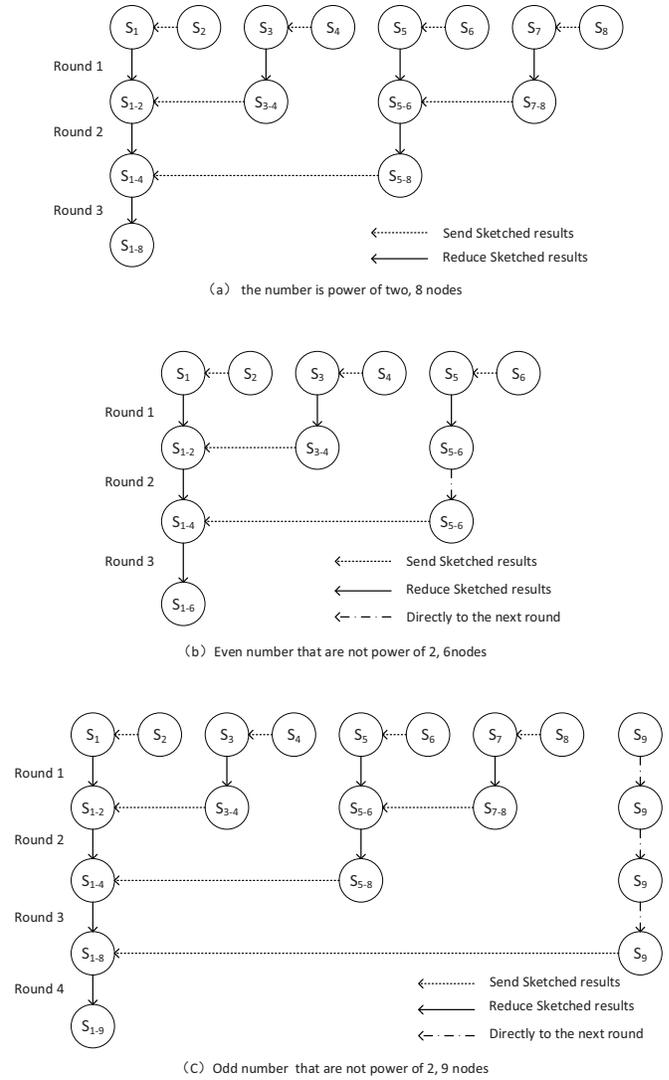}
	\caption{All-Reduce Process in gs-SGD when the number of workers is the power of 2 and non-power of 2}
	\label{fig:allreduce}
\end{figure}

% 介绍非2幂的情况
When $P$ is not a power of 2, Algorithm 1 needs some minor improvements. In Fig. \ref{fig:allreduce}(a), $P$ is power of two and Algorithm 1 works. In Fig. \ref{fig:allreduce}(b) and Fig. \ref{fig:allreduce}(c), $P$ is not power of two. From the initial round, when the number of nodes participating in the reduce is odd, the node with the largest id goes directly to the next round without participating in this round of recursive halving. Through mathematical induction, we can get that no matter the initial value is odd or even, it will eventually recurse to a unique root node. Therefore, no matter whether $P$ is a power of 2, we need at most $2\lceil \log P \rceil$ rounds to get all-reduce sketch results and pass them back to all workers.
% 介绍本课题中reduce的操作符含义
% 给出两张图，8和9的节点示意图，并总结分析算法复杂度
% 描述算法的过程

%gTopk 原文这里就是对取Top-k操作的定义，相关文章中也见到过这样的定义，我们的合并算子定义（Sketching Operation）：

% 下面这两个公式好像没有在文中引用，如果没用的话就删掉
% \begin{equation}
% 	\label{eq:hitter}
% 	\widetilde{G}^{a,b} = HEAVYMIX(S^{a,b})
% 	%其中
% \end{equation}

% \begin{equation}
% 	\label{eq:reduce}
% 	S^{a,b}=S^a+S^b= S(\widetilde{G}^{a})+S(\widetilde{G}^{b})
% \end{equation}

%就是先将本地梯度计算进Sketch中（$S^a,S^b,...$​​​），通过global AllReduce合并全部节点的sketch（$S^{a,b,...}$​​​），最后对这个总的sketch用HEAVYMIX算法进行取topk操作得到$\widetilde{G}^{a,b,...}$​​​。

% para2 g-top-k all reduce with sketch的详细介绍，配算法
% 介绍 T算子的定义等

% para1 方法的总体介绍，描述性介绍，分为几个步骤

% fig m1  拓扑结构的图

% non-power-of two

% para3 sketch  top k的计算复杂度分析
% sketch+topk的通信复杂度分析，P相关的，P假设为2的幂

% 收敛性的讨论

\floatname{algorithm}{Algorithm}
\renewcommand{\algorithmicrequire}{\textbf{Input:}}
\renewcommand{\algorithmicensure}{\textbf{Output:}}

\begin{algorithm}
	\caption{Global Sketching SGD}
	\begin{algorithmic}[1] %每行显示行号
		\Require The dataset: $D$
		\Require The initialized weights: $w$
		\Require The number of non-zero elements: $k$
		\Require The number of workers $P$
		\Require The rank of worker $g$
		\Require The learning rate $\eta $
		\Require The number of iterations to train: $T$
		%\Ensure $x$
		\State initialize the stochastic gradient $G^g_0 = 0$
		\For{$t = 1 \rightarrow T$}
		\State Sampling a mini-batch of data $D^g_t$ from $D;$
		\State Compute stochastic gradient $\widetilde{G}^g;$
		\State Compute sketches ${S}_t^g$ of $\widetilde{G}^g$
		\State $sends = {S}_t^g;$ 
		\State Initialize $recvs$ with the same as $sends;$
		\State $nRounds = \log P;$
		\For{$i = 1 \rightarrow nRounds$}
		\State $pRanks = [1 \rightarrow P, step = i];$
		\If {$i = 1 \rightarrow nRounds$}
		\State $localRank = pRanks.index(g);$
		\If {$localRank\% 2== 0$}
		\State $source = pRanks[localRank + 1];$
		\State Recv($recvs$, source=$source$)$;$
		\State $sends = recvs + sends;$
		\Else
		\State $target = pRanks[localRank - 1];$
		\State Send($sends$, dest=$target$)$;$
		\EndIf
		\EndIf
		\State $S_t = sends$ 
		\EndFor
		\State $\widetilde{G}_t = HEAVYMIX(S_t,k)$
		\State Update $w_{t+1} = w_t - \eta \widetilde{G}_t$
		\EndFor
	\end{algorithmic}
\end{algorithm}
	
\begin{algorithm}
	\caption{HEAVYMIX\cite{Ivkin2019CommunicationefficientDS}}
	\begin{algorithmic}[1] %每行显示行号
		\Require Sketch of gradient: $S$
		\Require The number of non-zero elements: $k$
		\Require The rank of worker: $g$
		\Ensure $\widetilde{g}:\forall{i}\in Top_k : \widetilde{g}_i = g_i$ and $\forall{i}\notin{Top_k}:\widetilde{g}_i = 0$
		
		\State $\forall{i}$ query $\hat{g}^2_i\pm\frac{1}{2k}||g||^2_2$ from sketch $S$
		
		\State $H \leftarrow {i|\hat{g}_i \geq \hat{l}^2_2/k}$ and $NH \leftarrow {i|\hat{g}_i < \hat{l}^2_2/k}$
		
		\State $Top_k = H\cup rand_l(NH),$ where $l = k - |H|$
		
		\State second round of communication to get exact values of Top$_k$
	\end{algorithmic}
\end{algorithm}

\subsection{Communication Complexity analysis}

%我们本文所讨论的通信节点均为通用GPU计算服务器，采用TCP/IP协议通信，因此具有同时接收和发送数据的能力。
%alpha beta 方法介绍
The communication nodes discussed in this article are all available GPU servers, which use the TCP/IP protocols to communicate to send and receive data simultaneously. For the communication complexity analysis, the cost of gradient exchanging between two nodes will be modeled by $\alpha + n \beta$ without considering network conflicts where $\alpha$ is message startup time, and $\beta$ is transmission time per data. Since all communication nodes are in the same data center, the communication time gap due to distance is negligible. In general, $\alpha$ is four to five orders of magnitude greater than $\beta$\cite{Chan2007CollectiveCT}. 

%分析我们算法的复杂度，每次sketch的复杂度，通信复杂度

In gs-SGD, the total all-reduce rounds are $\lceil\log P \rceil$, where $P$ is the number of workers. In each all-reduce round, gs-SGD guarantees that the number of workers participating is even. The length of sketched gradient vector has a complexity of $O(\log d)$ where $d$ is the length of gradient vector of deep learning models. Then, the total recursive halving and doubling time is Eq. \ref{eq:timecomplexity}

\begin{equation}
	\begin{aligned}
	t^{gs-SGD}_{all-reduce} &= 2 \lceil\log P \rceil (\alpha + \log d \cdot \beta) \\
	&= O(\log d \cdot \log P)
	\end{aligned}
\label{eq:timecomplexity}
\end{equation}

%\subsection{Convergence}

\section{Experimental Results}

%加一段整体实验部分的概述，要做哪些实验，用到什么环境，对比什么算法，验证哪些参数等
%%%修改
% 第一段要说一下，用什么模型，比较哪几种算法，加上这块
% 最后括号里的几个算法去掉吧，在正文里明确说）
% 全文的gs-SGD，统一，特别是实验部分的
We conduct extensive experiments to evaluate the effectiveness of our proposed gs-SGD by real-world applications on a 4-GPU cluster. We first compare the convergence efficiency with gs-SGD, gTop-$k$ S-SGD and Sketched-SGD on ResNet-20 and VGG-16. Then we evaluate and compare the time performance with gs-SGD, gTop-$k$ S-SGD and Sketched-SGD. After that, we run experiments on the convergence sensibility with different sparsity. We also make a comparison on scaling efficiency among the three algorithms.

%把下面的具体环境和对应的版本，列一个表
%实验部分的图顺序是乱的，按我和你说的，figure加label，然后ref的方法引用图编号，我上面有你可以参考。
%
All GPU machines are installed with Ubuntu-18.04, Nvidia GPU driver at version 460.56, and CUDA-11.2. The communication libraries are OpenMPI\cite{2008Open} at version 4.1.1 and NCCL \cite{nccl} at version 2.4. We use the highly optimized distributed training library Horovod at version 0.22.1. The deep learning framework is PyTorch at version 1.8.0 with cuDNN-11.1. Details of the software are shown in Table \ref{tab:software}.

\begin{table}[H]
	\caption{THE EXPERIMENTAL SETUP OF SOFTWARE}
	\centering
	\begin{tabular}{|c|l|}
		\hline 
		Software&Version\\
		\hline 
		OS&Ubuntu-18.04\\
		\hline 
		GPU driver&Nvidia GPU driver 460.56\\
		\hline 
		CUDA&11.2\\
		\hline 
		OpenMPI&4.1.1\\
		\hline 
		NCCL&2.4\\
		\hline 
		Horovod&0.22.1\\
		\hline 
		PyTorch&1.8.0\\
		\hline 
	\end{tabular}
	\label{tab:software}
\end{table}

\subsection{Convergence comparison}

% 添加了learning rates
We compare our gs-SGD with the gTop-$k$ S-SGD and the original Sketched-SGD with sparse gradients running on 4 workers. It should be noted that in order to improve the accuracy of our training, we used warmup for the first few epochs. To be specific, the first 4 epochs use the dynamic densities of [0.25, 0.0725, 0.015, 0.004] and small learning rates like [0.1, 0.03, 0.01].

\begin{figure}[H]
	\includegraphics[width=0.5\textwidth]{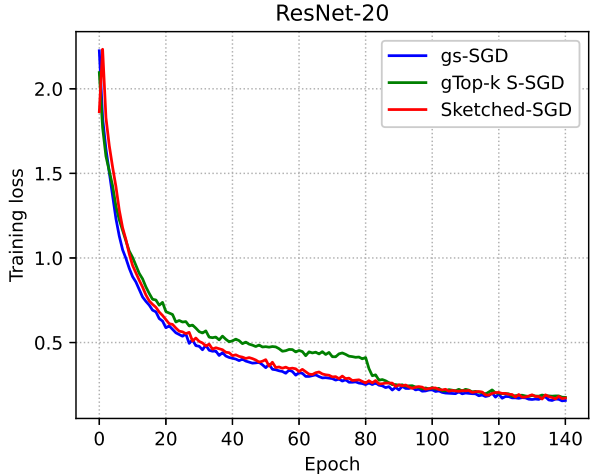}
	\caption{Convergence curves of ResNet-20 trained on Cifar-10 data set with gs-SGD, gTop-k S-SGD and Sketched-SGD with $P = 4$.}
	\label{fig:resnet_loss}
\end{figure}

\begin{figure}[H]
	\includegraphics[width=0.5\textwidth]{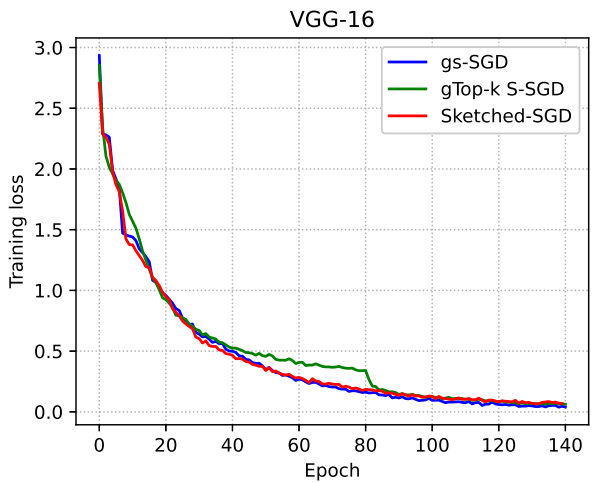}
	\caption{Convergence curves of VGG-16 trained on Cifar-10 data set with gs-SGD, gTop-k S-SGD and Sketched-SGD with $P = 4$.}
	\label{fig:vgg_loss}
\end{figure}

% 在85 epoch 之前，我们算法的收敛效率略好于gTopk S-SGD。
The Convergences of ResNet-20 and VGG-16 models on the Cifar-10 data set are shown in Fig. \ref{fig:resnet_loss} and Fig. \ref{fig:vgg_loss}. The results show that our gs-SGD basically converges at about 70 epochs, while gTop-$k$ needs about 85 epochs. In other words, the convergence efficiency of gs-SGD on ResNet-20 and VGG-16 is slightly better than that of gTop-$k$, especially before 85 epochs, and almost consistent with the convergence efficiency of original Sketched-SGD.

It can be seen that our algorithm has a better convergence rate than gTop-$k$ algorithm when training the above two models on the Cifar-10 data set, and does not lose the accuracy of the model.

\subsection{Time performance analysis}

\begin{figure}[H]
	\includegraphics[width=0.5\textwidth]{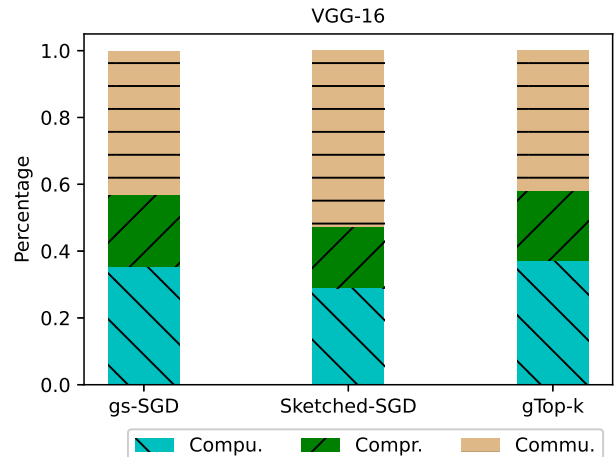}
	\caption{ Comparison of the time performance with gs-SGD, Sketched-SGD and gTop-$k$ on VGG-16 with $P=4$.}
	\label{fig:vgg_time_performance}
\end{figure}

\begin{figure}[H]
	\includegraphics[width=0.5\textwidth]{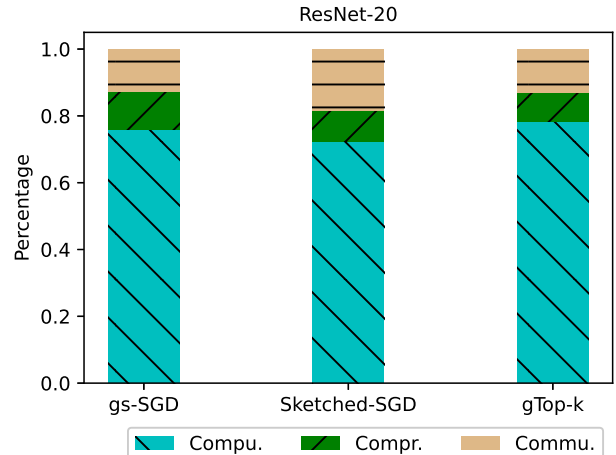}
	\caption{ Comparison of the time performance with gs-SGD, Sketched-SGD and gTop-$k$ on ResNet-20 with $P=4$.}
	\label{fig:resnet_time_performance}
\end{figure}

% 修改通信算法时间复杂度O(kp)和O(klogp)
% 我们的方法里，为什么还有k？应该是logd吧？
We use the cases of 4 workers to analyze the time performance of gs-SGD. We break down the time of an iteration into three parts: GPU computation time ($t_{compu.}$), local sparsification time ($t_{compr.}$), and communication time ($t_{commu.}$). The results are shown in Fig. \ref{fig:vgg_time_performance} and Fig. \ref{fig:resnet_time_performance}. It is obvious that the communication time of our algorithm is smaller than that of the original Sketched-SGD algorithm, especially on ResNet-20, because the AllReduce time complexity of our algorithm is $O(\log d*\log P)$ while the time complexity of the original Sketched-SGD algorithm is $O(\log d*P)$.

% 不讨论的话就删掉
\subsection{Convergence sensibility to the sparsity}

\begin{figure}[H]
	\includegraphics[width=0.5\textwidth]{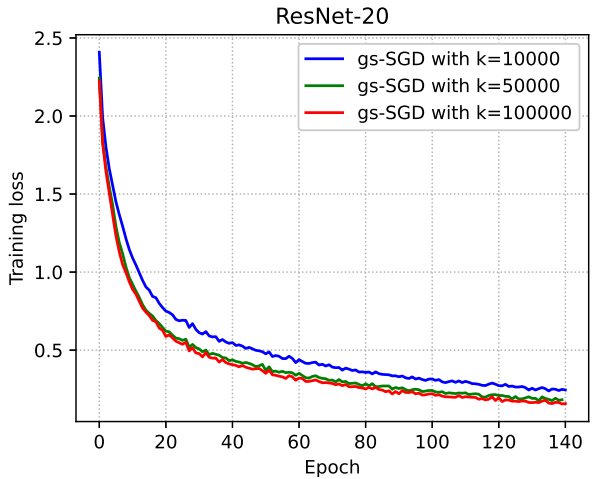}
	\caption{Convergence curves for ResNet-20 trained on Cifar-10 data set with different k on 4 workers.}
	\label{fig:resnet20_with_k}
\end{figure}

\begin{figure}[H]
	\includegraphics[width=0.5\textwidth]{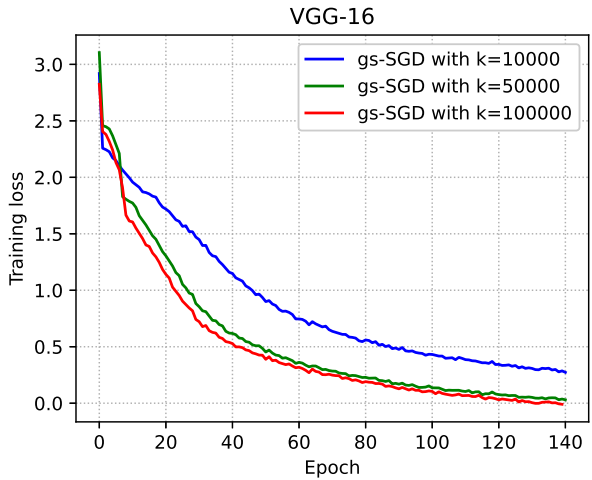}
	\caption{Convergence curves for VGG-16 trained on Cifar-10 data set with different k on 4 workers.}
	\label{fig:vgg_with_k}
\end{figure}

To test the sensibility of convergence of gs-SGD to the sparsity, we run the experiments with different numbers of the parameters $k$ using VGG-16 and ResNet-20 on the Cifar-10 data set on 4 workers. The convergence curves are shown in Fig. \ref{fig:resnet20_with_k} and Fig. \ref{fig:vgg_with_k}. It can be seen that a very small value of $k$-parameter of 10000 will significantly damage the convergence and accuracy of both models. Therefore, there is a tradeoff between the high compression ratio and a stable convergence speed. In other words, the higher compression ratio would bring lower communication overheads and higher scaling efficiency to a larger cluster. However, this indicates that we need to pay attention to the range of sparsity when training the model to avoid harming the convergence speed and accuracy of the model.

\subsection{Scaling efficiency}

In terms of scaling efficiency, we evaluate gs-SGD on a cluster with 4 GPU machines that are interconnected with 1 Gbps Ethernet. The summary of the training throughput on different models is shown in Table. \ref{tab:throughput}. The experimental results show that our method achieves 1.3—3.1× higher scaling efficiency than gTop-$k$ and 1.1—1.2× improvement than the Sketched-SGD. 

\begin{table}[H]
	\caption{THE SYSTEM TRAINING THROUGHPUT${^{\rm a}}$ ON A 4-GPU CLUSTER}
	\begin{threeparttable}
		{\begin{tabular}{cccccc}
			\toprule
			Model & gTop-$k$ & Sketched-SGD
					& gs-SGD
					& $g/k$$^{\rm b}$ & $g/s$$^{\rm c}$ \\
			\midrule
			ResNet-20 &804 & 976 & 1038 & 1.3× & 1.1× \\
			VGG-16 & 325 & 827 & 1014 & 3.1× & 1.2× \\
			\bottomrule
		\end{tabular}}
		\begin{tablenotes}
			\footnotesize
			\item[${\rm a}$] The throughput is measured with processed images per second.
			\item[${\rm b}$] indicates the speedup of gs-SGD compared to the gTop-$k$.
			\item[${\rm c}$] indicates the speedup of gs-SGD compared to Sketched-SGD.
		\end{tablenotes}
	\end{threeparttable}
	\label{tab:throughput}
\end{table}

\section{CONCLUSION}
% 梯度压缩是解决同步分布式优化中通信瓶颈的一种很有前途的方法。在这篇论文里，我们发现Sketch压缩方法的通信复杂度有O（kp），然而具有O（klogP）通信复杂度的gTop-k方法在训练模型时的收敛效率上还有提升空间。我们首次提出了Global-Sketch S-SGD，采用Count Sketch结构对梯度进行压缩和传输，在保证通信复杂度为O（klogP）的同时，提高了收敛效率和系统吞吐量。在不同卷积神经网络上的实验，包括VGG-16和ResNet-20，验证了Global-Sketch S-SGD对模型收敛性只是略有影响。我们在4-GPU的集群上进行了实验验证我们提出的Global-Sketch S-SGD比gtopk的收敛效率有所提升，并且在通信效率上略优于Sketched-SGD。
Gradient compression is a promising method to solve the communication bottleneck in synchronous distributed training. In this paper, we propose gs-SGD, which uses $Count-Sketch$ data structure to compress and transmit the gradients between workers, and improves convergence efficiency and system throughput while ensuring the communication complexity is O($\log d*\log P$). We conducted experiments to compare the convergence efficiency and time performance of gs-SGD, gTop-$k$ and Sketched-SGD. The experimental results show that our algorithm has significantly improved the convergence efficiency of training various CNNs, and reduces the communication time ratio among workers. Experiments on different convolutional neural networks, including VGG-16 and ResNet-20, verify that gs-SGD can guarantee convergence efficiency without losing model accuracy. We have conducted experiments on a 4-GPU cluster interconnected with 1Gbps Ethernet to confirm that the training throughput of gs-SGD proposed in this paper increased by 2.2× on average compared with gTop-$k$ S-SGD and 1.15× compared with Sketched-SGD.

For future work, we believe that scenarios with lower bandwidth network, more workers and more complex models are worth experimenting and researching.
% conference papers do not normally have an appendix

% use section* for acknowledgement
%\section*{Acknowledgment}
%The authors would like to thank...
%more thanks here

% trigger a \newpage just before the given reference
% number - used to balance the columns on the last page
% adjust value as needed - may need to be readjusted if
% the document is modified later
%\IEEEtriggeratref{8}
% The "triggered" command can be changed if desired:
%\IEEEtriggercmd{\enlargethispage{-5in}}

% references section

% can use a bibliography generated by BibTeX as a .bbl file
% BibTeX documentation can be easily obtained at:
% http://www.ctan.org/tex-archive/biblio/bibtex/contrib/doc/
% The IEEEtran BibTeX style support page is at:
% http://www.michaelshell.org/tex/ieeetran/bibtex/
\bibliographystyle{IEEEtran}
% argument is your BibTeX string definitions and bibliography database(s)
\bibliography{ref}
%
% <OR> manually copy in the resultant .bbl file
% set second argument of \begin to the number of references
% (used to reserve space for the reference number labels box)

% that's all folks
\end{document}